\documentclass[showpacs,amsmath,amssymb]{revtex4}

\usepackage{graphicx}
\usepackage{dcolumn}
\usepackage{bm}
\newcommand{\bnabla}{\mbox{\boldmath $\nabla$}}
\begin{document}

\preprint{SPIE 5514-45/Many-body interactions}

\title{Three- and four-body interactions in colloidal systems}

\author{Jure Dobnikar}
 \affiliation{University of Graz, Institute for Chemistry, Heinrichstrasse 28, 8010 Graz, Austria \\
 Institute Jo\v zef Stefan, Jamova 39, 1000 Ljubljana, Slovenia}
 \email{jure.dobnikar@uni-graz.at}
\author{Matthias Brunner}
 \affiliation{University of Stuttgart, 2. Physikalisches Institut, 70550 Stuttgart, Germany}
\author{J\" org Baumgartl}
 \affiliation{University of Stuttgart, 2. Physikalisches Institut, 70550 Stuttgart, Germany}
 \author{Hans-Hennig von Gr\" unberg}
 \affiliation{University of Graz, Institute for Chemistry, Heinrichstrasse 28, 8010 Graz, Austria }
\author{Clemens Bechinger}
 \affiliation{University of Stuttgart, 2. Physikalisches Institut, 70550 Stuttgart, Germany}

\date{\today}

\begin{abstract}
  Three-body and four-body interactions have been directly measured
  in a colloidal system comprised of three (or four) charged colloidal
  particles. Two of the particles have been confined by means of a
  scanned laser tweezers to a line-shaped optical trap where they
  diffused due to thermal fluctuations. By means of an additional 
  focused optical trap a third particle has been approached and 
  attractive three-body interactions have been observed. These observations 
  are in qualitative agreement with additionally performed nonlinear 
  Poissson-Boltzmann calculations. Two configurations of four particles
  have been studied experimentally as well and in both cases a repulsive
  four-body interaction term has been observed.
\end{abstract}

\pacs{Valid PACS appear here}

\maketitle
 
\section{Introduction}

The total interaction energy of a system comprised of $N$ particles consists of 
pair terms, three-body, four-body, ... and $N$-body contributions. The complete 
description of such a system should, in principle, take into account all
these terms. In practice, however, the $N$-body terms are rarely known, and 
even if they are, it is a formidable task to include even a few lower order 
terms in any physical theory. So, in most cases, the description is grossly 
simplified by approximating the complete many-body description of the system 
by {\sl effective} pair interactions. All higher many-body terms are 
integrated out into the effective parameters appearing in the effective pair 
interactions. Due to the pairwise additivity of the simplified model, the 
evaluation of physical properties of the system becomes tractable, but one 
should be aware of the following limitations.

The effective pair interactions are highly integrated quantities and depend
on the particular many-body configuration of the system. Therefore, there is no
unique effective pair interaction describing a system. Moreover, the 
simplification very often leads to thermodynamic inconsistencies \cite{Louis}. 
The effective quantities depend on the state of the system, i.e. density, 
and even on the arrangement of the particles in space. This is illustrated in 
\cite{EPL,JCP,gr} where the effective pair interactions are determined in a solution 
of charged colloids. In \cite{gr} the effective interactions are extracted from
radial distribution functions measured in a 2D colloidal suspension at different 
densities. In those experiments a clear dependence of the results on the density 
of colloids has been observed. In \cite{EPL,JCP,JPCM} the effective pair interactions 
have been studied numerically in a 3D colloidal suspension, solving the non-linear 
Poisson-Boltzmann equation and integrating out the many-body terms. The resulting 
effective pair interaction was different when the colloids were placed in a FCC 
lattice than when they were forming a BCC lattice.

These examples illustrate that the effective potentials extracted from experimental 
studies by simply determining the effective parameters to achieve agreement 
between theory and experiment may well work, but as we have seen, it is not a 
fundamental approach and has severe limits of applicability. Any correct physical 
description of a many-body system should therefore take many-body terms into account.

Already in 1943 it has been supposed by Axilrod and Teller (AT)
\cite{Teller} and later also by Barker and Henderson \cite{Barker}
that three-body interactions may significantly contribute to the total
interaction energy in noble gas systems. This seems to be surprising
because noble gas atoms posses a closed-shell electronic structure and
are therefore often (and erroneously) regarded as an example of a
simple liquid . The conjecture of Axilrod and Teller, however, was
confirmed only very recently, when large-scale molecular dynamics
simulations for liquid xenon and krypton \cite{Bomont,Jakse} was
compared with structure factor measurements at small q-vectors
performed with small-angle neutron scattering
\cite{Formisano1,Formisano2}. In these papers it has clearly been
demonstrated that only a combination of pair-potentials and three-body
interactions, the latter in the form of the AT-triple-dipole term
\cite{Teller}, leads to a satisfactory agreement with the experimental
data.  In the meantime, it has been realized that many-body
interactions have to be considered also for nuclear interactions
\cite{Negele}, inter atomic potentials, electron screening in metals
\cite{Hafner}, photo-ionization, island distribution on surfaces
\cite{Osterlund}, and even for the simplest chemical processes in
solids \cite{Ovchinnikov} like breaking or making of a bond.

In view of the general importance of many-body effects it seems
surprising that until now no direct measurements of these interactions
have been performed. This is largely due to the fact that in atomic
systems, positional information is typically provided by structure
factors or pair-correlation functions, i.e. in an integrated form.
Direct measurements of many-body interactions, however, require direct
positional information beyond the level of pair-correlations, which is
not accessible in atomic or nuclear systems. In contrast to that,
owing to the convenient time and length scales involved, the
microscopic information is directly accessible in colloidal
suspensions. In addition, the pair-interactions in colloidal
suspensions can be varied over large ranges, e.g. from short-ranged
steric to long-ranged electrostatic or even dipole-dipole
interactions.

In the present study \cite{PRL,PRE} we used charged colloidal particles whose
interactions are mediated by the microscopic ions in the
electrolyte. The pair interaction in such systems is directly related
to the overlap of the ion clouds (double-layers) which form around the
individual colloids and whose thickness is determined by the ionic
strength of the solution. In highly de-ionized solutions, these
double-layers can extend over considerable distances. If more than two
colloids are within the range of such an extended double layer, many-body 
interactions are inevitably the consequence. Accordingly, deviations from 
pair-wise additive interactions are expected under low salt conditions and
high densities of colloids.

Experimental evidence for many-body interactions has been already
obtained from effective pair-interaction potential measurements of
two-dimensional (2D) colloidal systems. Upon variation of the particle
density, a characteristic dependence of the effective pair interaction
was found which has been interpreted in terms of many-body
interactions \cite{gr}. However, during those studies the relative
contributions of different many-body terms could not be further
resolved. Performing the experiment described in this paper, i.e.
observing the system of only three (or four) particles, we were able 
to measure the three- and four-body interactions directly. 

\section{Experimental system}

As colloidal particles we used charge-stabilized silica spheres with
990nm diameter suspended in water. A highly diluted suspension was
confined in a silica glass cuvette with 200$\mu$m spacing. The cuvette
was connected to a closed circuit, to deionize the suspension and thus
to increase the interaction range between the spheres. This circuit
consisted of the sample cell, an electrical conductivity meter, a
vessel of ion exchange resin, a reservoir basin and a peristaltic pump
\cite{Palberg}. Before each measurement the water was pumped through
the ion exchanger and typical ionic conductivities below 0.07$\mu$S/cm
were obtained. Afterwards a highly diluted colloidal suspension was
injected into the cell, which was then disconnected from the circuit
during the measurements. This procedure yielded stable and
reproducible ionic conditions during the experiments. Due to the ion
diffusion into the sample cell, the screening length $\kappa^{-1}$
decreased linearly with time during the measurements. The rate of
change of the screening length, however, was only less than half a
percent per hour, which means that in the time needed to perform a
complete set of measurements, the ionic concentration did not change
more than about 1 percent. This tiny variation has been taken into
account when performing the PB calculations (see section IV.).\\

\begin{figure}
 \includegraphics[width=0.75\textwidth]{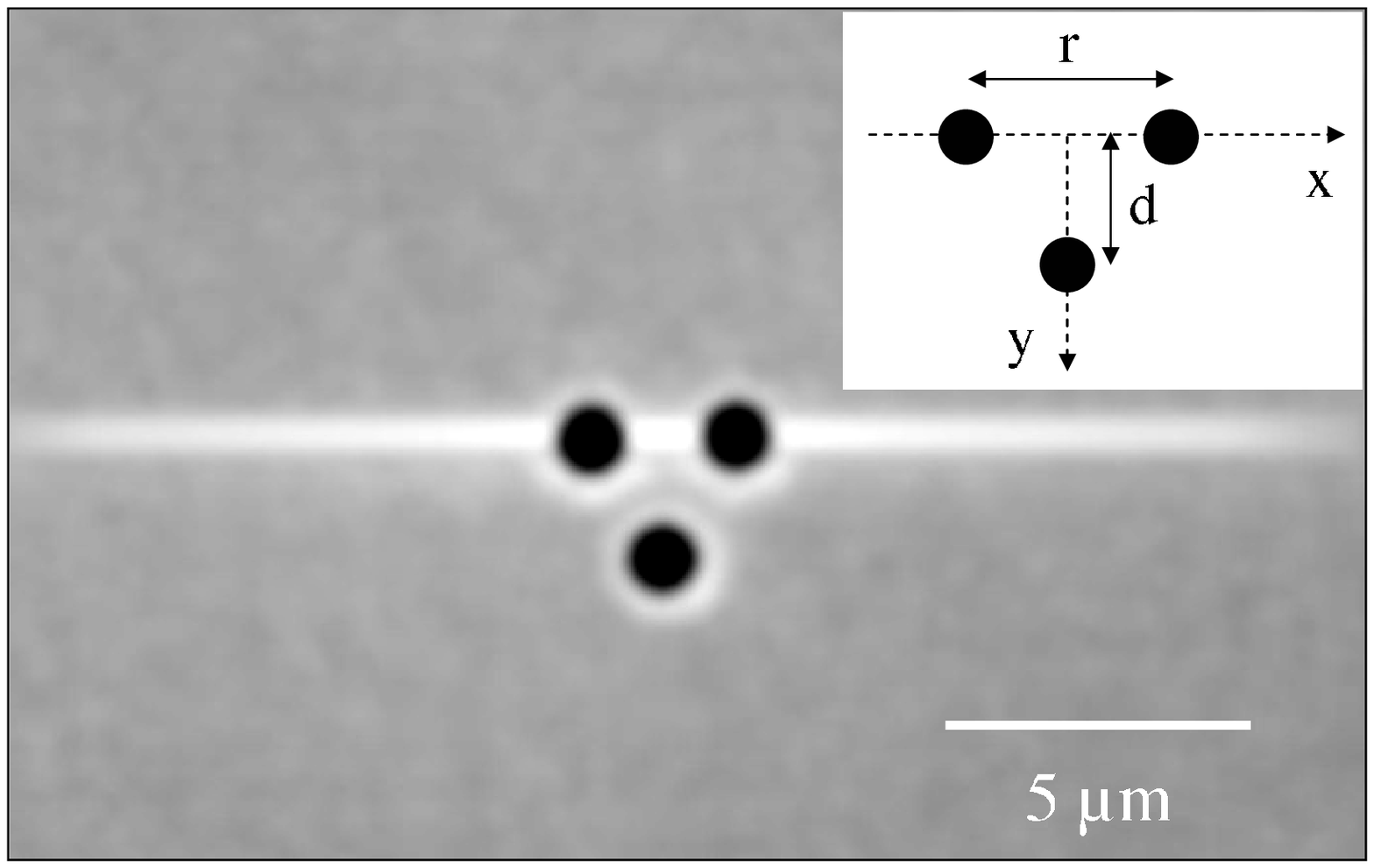}
 \caption{Photograph of sample cell with two silica particles
confined to a light trap created by an optical tweezers and a third
particle trapped in a focused laser beam. The inset shows a schematic
drawing of the experimental geometry.}
\label{photo}
\end{figure}
After the particles sedimented down to the bottom plate of the
sample cell, they were brought in the field of view of the
microscope (Fig.\ref{photo}). Two particles were trapped with
line-scanned optical tweezers, which was created by the beam of an
argon ion laser being deflected by a computer-controlled
galvanostatically driven mirror with a frequency of approximately
350Hz.  The time averaged intensity along the scanned line was chosen
to be Gaussian-distributed with the half-width $\sigma_{x}\approx$ 4.5
$\mu$m. The laser intensity distribution perpendicular to the trap was
given by the spot size of the laser focus, which is also Gaussian with
$\sigma_{y}\approx$ 0.5 $\mu$m. This yielded an external laser
potential acting as a stable quasistatic trap for the particles.  Due
to the negatively charged silica substrate, the particles also
experience a repulsive vertical force, which is balanced by the
particle weight and the vertical component of the light force. Due 
to the fact that the potential is much steeper in the vertical 
direction than the in the $xy$ plane, the vertical particle 
fluctuations can be disregarded. The particles were imaged with 
a long-distance, high numerical aperture microscope objective 
(magnification $\times$63) onto a CCD camera and the images were 
stored every 120 ms.  The lateral positions of the particle centers 
were determined with a resolution of about 25 nm by a particle 
recognition algorithm.

Three- and four-body interaction potentials were measured in this 
setup by performing the following steps (which will be explained in 
detail below): First only one particle was inserted into the trap and its
position probability distribution was evaluated from the recorded
positions. From this the external laser potential $u_{L}$ could be
extracted. Next, we inserted two particles in the trap and measured
their distance distribution. From this, the pair-interaction potential
was obtained. Finally, a third particle was made to approach to the
optical trap by means of additional point optical tweezers (focus size
$\approx$ 1.3 $\mu$m), which held this particle at a fixed position
during the measurement. From the distance distribution of the first
two particles we obtained the total interaction potential for the
three particles. Finally, we substracted a superposition of pair
potentials (known from the previous two-particle measurements) from
the total interaction energy to obtain the three-body interaction. The 
same procedure was done with four particles to determine the four-body 
interaction. The fourth particle was fixed with additional point laser 
tweezers. The many-body term obtained after subtracting the pair terms 
was now the sum of three- and four-body interactions.

\section{Data evaluation and experimental results}

\begin{figure}
 \includegraphics[width=0.75\textwidth]{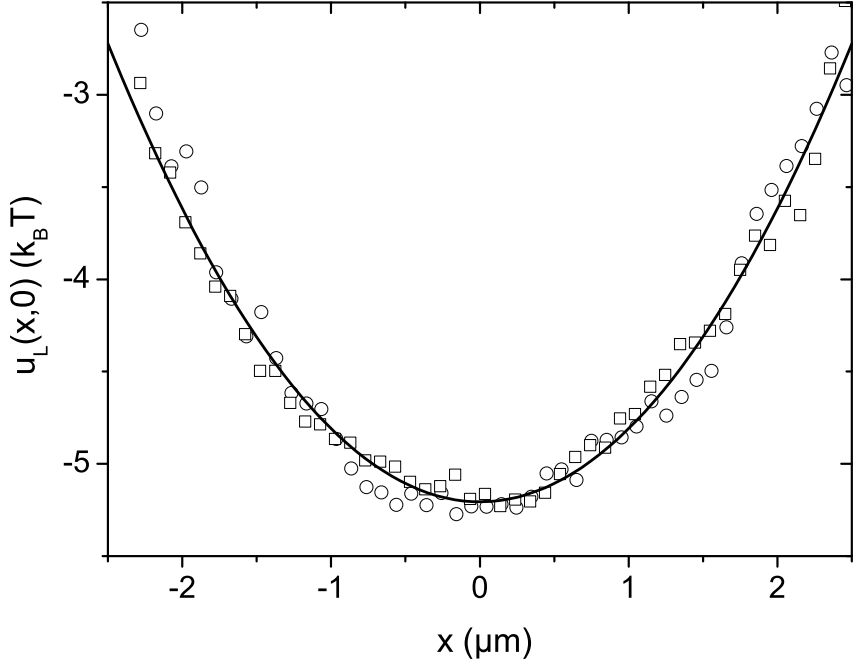}
 \caption{The shape of the laser potential along the tweezers line
for three different laser intensities (symbols: triangles 100mW,
circles 200mW and squares 500mW); for better comparison all curves are
normalized to an intensity of 100mW. The Gaussian fit is plotted as a
solid line.}
\label{laserpot}
\end{figure}

We first determined the external potential acting on a single particle
due to the optical line trap. The probability distribution $P(x,y)$ of
finding a particle at the position $(x,y)$ in the trap was evaluated
from the recorded positions. $P(x,y)$ depends only on the temperature
and the external potential $u_{L}(x,y)$ created by the laser tweezers.
According to the Boltzmann probability distribution
$P(x,y)=P_{L}\rm{e}^{-\beta u_{L}(x,y)}$, with $P_{L}$ being a
normalization constant. Taking the logarithm of $P(x,y)$ yields the
external potential $u_{L}(x,y)$ with an offset given by $\log{P_{L}}$.
The probability distributions in $x$ and $y$ directions are
statistically independent, and can therefore be factorized. The laser
potential is thus $u_{L}(x,y)=u_{L}(x)+u_{L}(y)$. The potential along
the $x$ axis is shown in Fig.\ref{laserpot} for various laser
intensities.  
As can be seen, all renormalized potentials fall, within
our experimental resolution, on top of each other. This clearly
demonstrates that the optical forces exerted on the particles scale
linearly with the input laser intensity. This fact allows us to use
different external laser powers for two-, three- and four-body experiments
(in the three- and four-body experiment, due to the additional repulsion 
of the third particle, a stronger laser power is needed to keep the mean
distance between the two particles similar). The corresponding
potential in the perpendicular ($y$) direction has the same (Gaussian)
shape, but it is much steeper due to the chosen scanning direction.
Therefore, the particles hardly move in the $y$ direction during a
measurement.

\begin{figure}
 \includegraphics[width=0.75\textwidth]{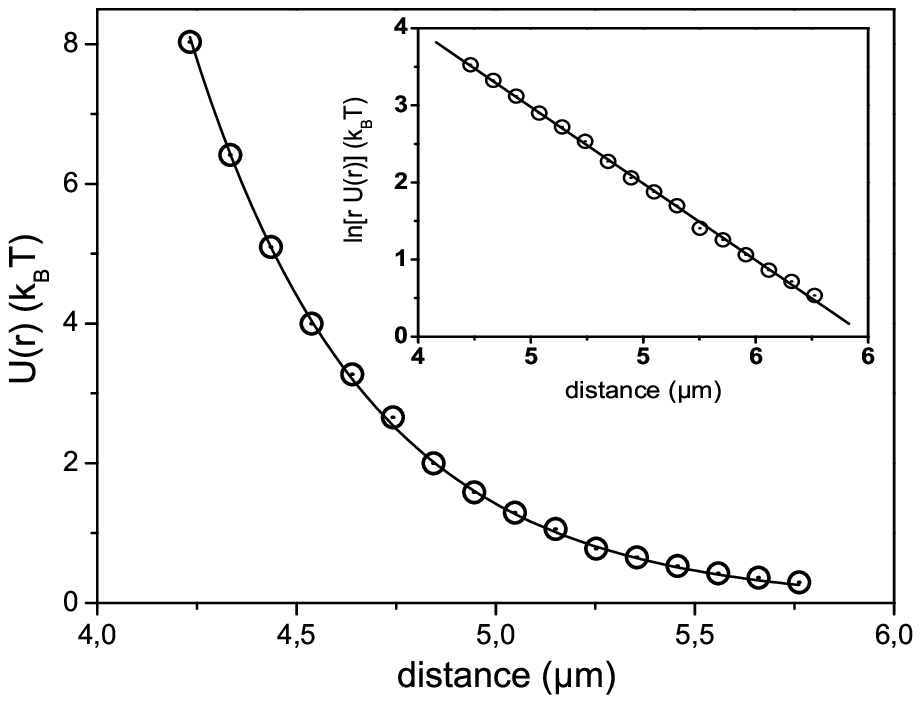}
 \caption{Measured pair-interaction potentials $U(r)=u_{pair}(r)$
(symbols) in the absence of the third particle.  The data agree well
with a DLVO potential, Eq.(\ref{Yukawa}) (solid line). In the inset
the potential is multiplied by $r$ and plotted logarithmically, so
that the DLVO expression, Eq.(\ref{Yukawa}) transforms into a straight
line. From a fit we obtained the effective charge $Z^{*}\approx 6500$
and the screening length $\kappa^{-1} \approx
470$nm.}
\label{pairpot}
\end{figure}
Next, we inserted a second particle in the trap. The four-dimensional
probability distribution is now
$P(x_{1},y_{1},x_{2},y_{2})=P_{12}\rm{e}^{-\beta \left
(u_{L}(x_{1},y_{1})+u_{L}(x_{2},y_{2})+U(r)\right )}$, with $x_{i}$
and $y_{i}$ being the positions of the $i$-th particle relative to the
laser potential minimum and $U(r)$ the distance dependent
pair-interaction potential between the particles. This can be
projected to
\begin{widetext}
\begin{eqnarray}
P(r)&=&\int\!\!\!\int\!\!\!\int\!\!\!\int P(x_{1},y_{1},x_{2},y_{2})
\delta\!\left (\!\sqrt{(x_{1}-x_{2})^2 + (y_{1}-y_{2})^2}-r\right )
dx_{1}dx_{2}dy_{1}dy_{2} = 
\nonumber \\
&=&P_{12}\rm{e}^{-\beta U(r)}\!\int\!\!\!\int\!\!\!\int\!\!\!\int 
\rm{e}^{-\beta \left (u_{L}(x_{1},y_{1})+u_{L}(x_{2},y_{2})\right )}
 \, \delta\!\left (\!\sqrt{(x_{1}-x_{2})^2 + (y_{1}-y_{2})^2}-r\right ) 
dx_{1}dx_{2}dy_{1}dy_{2}
\label{projection}
\end{eqnarray}
\end{widetext}
In principle the integral is constituted of all possible
configurations of two particles with distance $r$. Performing the full
four-dimensional integration, however, is difficult because of the
limited experimental statistics. This problem can be overcome by the
following two considerations. First, due to the Gaussian shape of the
external potential, the most likely particle configurations are
symmetric with respect to the potential minimum of $u_{L}$ (any
asymmetric configuration for constant $r$ has a higher
energy). Secondly, particle displacements in $y$-direction are
energetically unfavorable because $\sigma_{x} \gg
\sigma_{y}$. Accordingly, for $r = const$ the minimum energy
configuration is $(x_{1}=r/2, y_{1}=0, x_{2}= -r/2, y_{2}=0)$. It has
been confirmed by a simple calculation with the experimental
parameters that all other configurations account for only for less than 1
percent of the value of the integral in
Eq.(\ref{projection}). Accordingly, Eq.(\ref{projection}) reduces to
\begin{eqnarray}
P(r)=P_{0}\rm{e}^{-\beta \left ( U(r)+2u_{L}(r/2,0) \right )} \; .
\label{Pr}
\end{eqnarray}
Since $u_{L}(x,y)$ is known from the previous one-colloid measurement,
we can obtain the interaction potential $U(r)$ from the measured $P(r)$,
\begin{eqnarray}
\beta U(r)\, =\, -\log{P(r)}\, -\, 2\beta u_{L}(r/2,0)\, +\, \log{P_{0}}\;.
\label{Ur}
\end{eqnarray}
The normalization constant $P_{0}$ was chosen in a way that $U(r) \to
0$ for large particle separations $r$.  We first measured $U(r)$
according to the above procedure in the absence of a third
particle. As expected, the negatively charged colloids experience a
strong electrostatic repulsion which increases with decreasing
distance. The pair-interaction potential of two charged spherical
particles in the bulk is well known to be described by a Yukawa
potential \cite{Landau,Overbeek}
\begin{eqnarray}
\beta U(r)=\beta u_{pair}(r)=\left ( Z^{*}\right ) ^2
\lambda_{B} \Biggl (\frac{\rm{e}^{\kappa R}}{1+\kappa R}\Biggr )^2
\frac{\rm{e}^{-\kappa r}}{r} \; ,
\label{Yukawa}
\end{eqnarray}
where $Z^{*}$ is the renormalized charge \cite{Belloni} of the
particles, $\lambda_{B}$ the Bjerrum-length characterizing the solvent
($\lambda_{B}=e^2/4\pi\epsilon\epsilon_{0}k_{B}T$, with $\epsilon$ the
dielectric constant of the solvent and $e$ the elementary charge),
$\kappa^{-1}$ the Debye screening length (given by the salt
concentration in the solution), $R$ the particle radius and $r$ the
centre-centre distance of the particles. Fig.\ref{pairpot} shows the
experimentally determined pair-potential (symbols) together with a fit
to Eq.(\ref{Yukawa}) (solid line). As can be seen, our data are well
described by Eq.(\ref{Yukawa}). As fitting parameters we obtained
$Z^{*}\approx 6500$ electron charges and $\kappa^{-1} \approx 470$nm,
respectively. The renormalized charge is in good agreement with the
predicted value of the saturated effective charge of our particles
\cite{Zsat,Trizac} and the screening length agrees reasonably with the
bulk salt concentration in our suspension as obtained from the ionic
conductivity. Given the additional presence of a charged substrate, it
might seem surprising that Eq.(\ref{Yukawa}) describes our data
successfully. However, it has been demonstrated experimentally
\cite{Grier} and theoretically \cite{Stillinger,Netz} that a
Yukawa-potential captures the leading order interaction also for
colloids close to a charged wall. A confining wall introduces only a
very weak (below 0.1 $k_{B}T$) correction due to additional dipole
repulsion. This correction is below our experimental resolution.
Repeating the two-body measurements with different laser intensities
(50mW to 600mW) yielded within our experimental resolution identical
pair potential parameters. This also demonstrates that possible
light-induced particle interactions (e.g. optical binding
\cite{Burns}) are neglegible. The approach of the third and fourth 
particle by means of an additional optical trap could, in principle, 
lead to additional light-induced interactions between the laser spot 
and the two particles kept in the line trap. To exclude such effects, we
repeated the two-particle measurements and approached an empty trap
(without the third particle) to the line trap where the two particles
were fluctuating. Within our experimental resolution, we again
observed identical pair potentials, which suggests, that the
additional optical trap has no influence on the two particles in the
line trap.  When the third or fourth particle are present at a distance 
$d$ along the perpendicular bisector of the scanned laser line (cf. inset of
Fig.\ref{photo}), the total interaction energy $U(r,d)$ is not simply
given by the sum of the pair-interaction potentials Eq.(\ref{Yukawa})
alone but also contains additional many-body terms. Following the 
definition of McMillan and Mayer \cite{McM}, $U(r,d)$ is given by
\begin{figure}
 \includegraphics[width=0.75\textwidth]{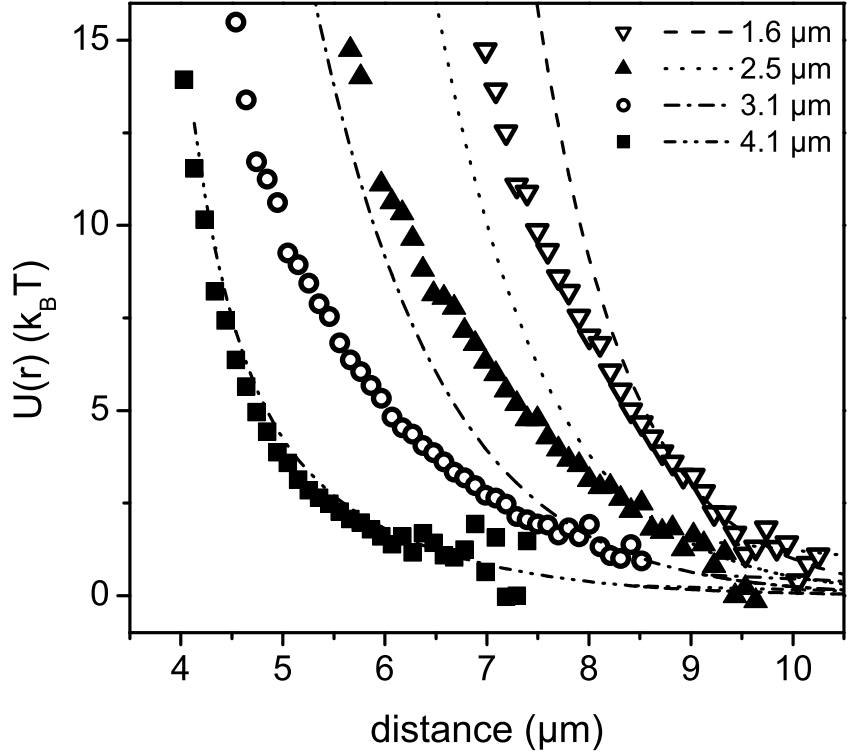}
 \caption{Experimentally determined interaction energy $U(r)$
(symbols) for two particles in a line tweezers in the presence of a
fixed third particle with distance $d$ on the perpendicular bisector of
the line trap. For comparison the superposition of three
pair-potentials is plotted as lines. Symbols and lines are labelled by
the value of $d$.}
\label{utot}
\end{figure}
\begin{eqnarray}
U(r,d) = \sum_{i\ne j}u_{pair} (r_{ij}) + u_{\rm many-body}\; , 
\label{uMB}
\end{eqnarray}
with $u_{pair}(r_{ij})$ being the pair-potential between particles $i$
and $j$ as defined in Eq.(\ref{Yukawa}), indices $i$ and $j$ run from 
1 to the number of particles. The last term,  $u_{\rm many-body}$ is 
the sum of many-body interaction potentials. In the case of three particles
$u_{\rm many-body} \equiv u_{123}(r_{12},r_{13},r_{23})$ and Eq.(\ref{uMB}) 
reduces to 
\begin{widetext}
\begin{eqnarray}
U(r,d) = u_{pair} (r_{12}) + u_{pair} (r_{13}) + u_{pair} (r_{23}) + 
u_{123}(r_{12},r_{13},r_{23})\; . 
\label{u123}
\end{eqnarray}
\end{widetext}
Distances $r_{ij}$  are the distances between the three particles in the system which can, 
due to the chosen symmetric configuration ($r_{23}\equiv r_{13}$), be expressed by the
two variables $r = r_{12}$ and $d=\sqrt{r_{13}^{2}-(r/2)^2}$. We have
followed the same procedure as described above for the case of two
particles. First, we have measured the probability distribution
$P(r;d)$ of the two particles in the laser trap with the third
particle fixed at distance $d$ from the trap.  Taking the logarithm of
$P(r;d)$ we extracted the total interaction energy $U(r,d)$.
When approaching the third particle, the two particles in the trap are
slightly displaced in the $y$-direction at small $r$. Accordingly, the
minimum energy configurations of the two particles are not on a
straight line as before. The most likely configuration at given
distance $r$ is $\left (x_{1}=r/2, y_{1}=y(r), x_{2}=-r/2,
y_{2}=y(r)\right )$ with $y(r)$ given by the measured particle
positions. Since we have the full knowledge of the two-dimensional
external laser potential, we can compute $u_{L}(r/2,y(r))$ for every
given configuration and use it in the Eq.(\ref{Pr}) instead of
$u_{L}(r/2,0)$. The results are plotted as symbols in Fig.\ref{utot} 
for the distance of the third particle $d = 4.1,$ 3.1, 2.5 and 1.6 $\mu$m,
respectively. As expected, $U(r,d)$ becomes larger as $d$ decreases
due to the additional repulsion between the two particles in the trap
and the third particle. In order to test whether the interaction
potential can be understood in terms of a pure superposition of
pair-interactions, we first calculated $U(r,d)$ according to
Eq.(\ref{u123}) with $u_{123} \equiv 0$. This was easily achieved
because the positions of all three particles were determined during
the experiment and the distance-dependent pair-potential is known from
the two-particle measurement described above (Fig.\ref{pairpot}). The
results are plotted as dashed lines in Fig.\ref{utot}. Considerable deviations
from the experimental data can be observed, in particular at smaller
$d$. These deviations can only be explained, if we take three-body
interactions into account. Obviously, at the largest distance, i.e. $d
= 4.1 \mu m$ our data are well described by a sum over pair-potentials
which is not surprising, since the third particle cannot influence the
interaction between the other two, if it is far away from both. In
agreement with theoretical predictions \cite{Carsten}, the three-body
interactions therefore decrease with increasing distance $d$.

\begin{figure}
 \includegraphics[width=0.75\textwidth]{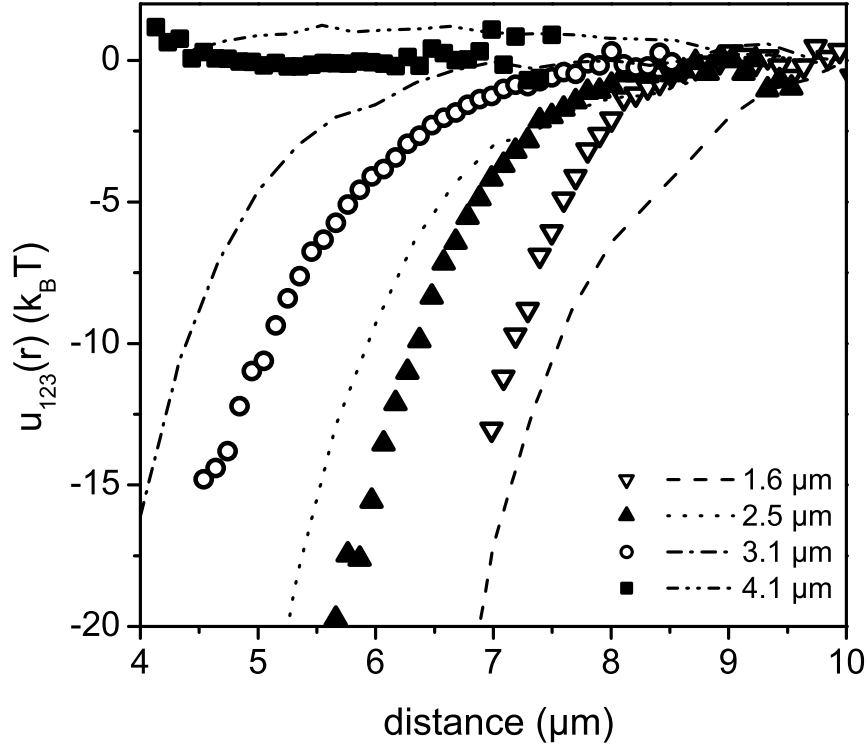}
 \caption{Three-body potentials for different $d$. Measured
three-body potentials indicated by symbols. The lines are three-body
potentials as obtained from the solutions of the nonlinear
Poisson-Boltzmann equation for three colloids arranged as in the
experiment. The parameters in the Poisson-Boltzmann calculation were
chosen so that the pair-interaction potentials were correctly
reproduced. Symbols and lines are labeled by the value of $d$.}
\label{u3}
\end{figure}
According to Eq.(\ref{u123}) the three-body interaction potential 
(the only many-body term in case of only three particles) is
simply given by the difference between the measured $U(r,d)$ and the
sum of the pair-potentials (i.e. by the difference between the
measured data and their corresponding lines in Fig.\ref{utot}). The
results are plotted as symbols in Fig.\ref{u3}. It is clearly seen,
that in the case of charged colloids $u_{123}$ is entirely attractive
and becomes stronger as the third particle approaches. It is also
interesting to see that the range of $u_{123}$ is of the same order as
the pair-interaction potentials. It might seem surprising that it is
possible to sample the potential up to energies of 15$k_{B}T$, as
configurations of such a high energy statistically happen only with
very low probability. In this experiment we can choose the energetic
range of the potential we want to sample by adjusting the strength of
the line tweezers. The laser potential pushes the particles together,
which allows us to sample different ranges of the electrostatic
potential. Thus, to achieve a better resolution for smaller particle
separations (e.g. higher potential values), the strength of the line
tweezers had to be increased. The shape of the external potential
$u_{L}$ was independent of the strength of the laser beam (see
Fig.\ref{laserpot}) and the magnitude scaled linearly with the input
laser power. This allowed us to adjust the input laser intensity so as
to obtain a suitable particle separation range. The external potential
was obtained simply by scaling the Gaussian shown in Fig.\ref{laserpot}.

\section{Four-body interactions}

The fourth particle has been placed symmetrically to the third one at the
distance $d$ from the line trap by means of an additional focussed 
laser tweezers. Like the third particle it was kept fixed at this position 
during the experiment. The data evaluation is identical to the evaluation
in the case of three particles, the distance probability $P(r)$ of the two 
particles in the line trap is analysed, from which the total interaction 
potential is obtained. After subtracting the pair energies we have obtained 
the total many-body interaction potential $u_{\rm many-body}$. In the case of 
three particles the total many-body interaction potential was simply the 
three-body term $u_{123}$, but with four particles it consists of more terms.
In the experiment the interaction between the two colloids in the line trap was 
observed. When the third particle was introduced, the interaction was altered
and the amount to which it has changed was called the three-body interaction 
potential. Now, as we add another particle, we expect to see the same three-body
effect once again, i.e. the total change of interaction should be twice the 
three-body term.This would be true if there were no four-body interactions, but 
with the four-body term the total many-body interaction potential is 
$u_{\rm many-body}=2u_{123}(r,d) + u_{1234}(r,d)$. The four-body interaction 
potential can thus be extracted,
 
\begin{eqnarray}
 u_{1234}(r,d)=U(r,d)-\sum_{i\ne j} u_{pair} (r_{ij}) - 2u_{123}(r,d) \; .  
\label{u4}
\end{eqnarray}

\begin{figure*}
 \includegraphics[angle=-90,width=0.75\textwidth]{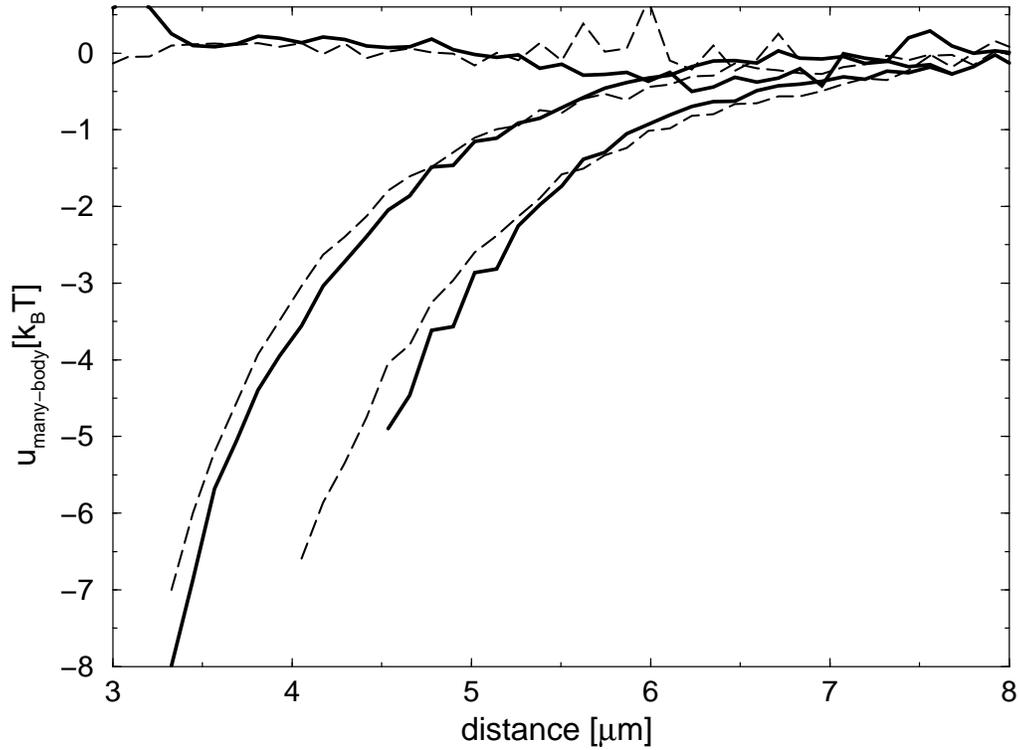}
 \caption{Experimentally determined total many-body potential 
(three- and four-body terms superposed) in the configuration of four 
particles (solid lines) plotted together with the three-body potential in 
the three particle case (dashed lines). The three sets of curves correspond 
to different distances $d$ of the third and the fourth particle from the 
baseline. From bellow they correspond to $d=2.4\mu m$, $d=3.0\mu m$ and 
$d=4.2\mu m$. In the latter case the third and fourth particles are so 
far away, that the measured many-body interaction vanishes. In the other 
two cases the four-body interaction term is opposite and of the same size 
as the three-body term, so that the total many-body interaction potential, 
which is $u_{\rm many-body}=2u_{123}+u_{1224}$, looks almost as the three-body
term. See also the text in the section {\sl Four-body interactions}. The 
three-body interaction potential here differs from the one on Fig.\ref{u3}, 
while the experimental conditions (salt concentration) were slightly 
different in this set of measurements.}
\label{u4b}
\end{figure*}

The four-body potential $u_{1234}(r,d)$ depends only on the two 
parameters $(r,d)$ due to the chosen symmetric configuration. The pair and 
three-body terms are obtained from the previous measurements. The results 
are presented on Fig.\ref{u4b}. The solid lines represent the many-body 
interaction energies obtained from the four-body experiments. The dashed lines 
are the three-body interaction energies obtained from the three-body experiments 
at the same conditions. There are three pairs of curves on the figure. 
Each pair corresponds to a certain distance $d$ of the third and fourth particle 
from the line trap. From the upper to the lower pair the distances were 
$d=4.2\mu m, 3.0\mu m, 2.4\mu m$. Again, when the third and fourth particles were very 
far from the line trap $(d=4.2\mu m)$, the many-body term vanishes. When they are 
placed closer, there is an attractive three-body interaction already discussed 
(dashed lines, $d=3.0\mu m, 2.4\mu m$). Also the total many-body interaction in
the four-body experiment (solid lines, $d=3.0\mu m, 2.4\mu m$) is attractive and, 
curiously, quite similar to the three-body term. Applying Eq.\ref{u4} we can conclude 
that the four-body interaction is repulsive and of about the same range and magnitude 
as the attractive three-body term,

\begin{eqnarray}
 u_{1234}(r,d) \approx - u_{123}(r,d) \; .  
\label{u4u3}
\end{eqnarray}

The curiosity appearing in the two configurations explored, is that the total many-body 
interaction potential lies almost on top of the three-body term. If this was a 
universal conclusion, it would mean that the three-body terms are partially canceled by
the four-body interaction potentials, i.e., we could include the four-body interaction 
simply by counting only half of the three-body terms. However, while this would certainly 
be an interesting feature, we must be aware of the fact that only two single 
configurations were studied here, not enabling us to draw very general conclusions. 

\section{Numerical calculations}

In order to get more information about three-body potentials in
colloidal systems, we additionally performed non-linear
Poisson-Boltzmann (PB) calculations \cite{CPC}, in a similar way as in
\cite{Carsten}. The PB theory provides a mean-field description in
which the micro-ions in the solvent are treated within a continuum
approach, neglecting correlation effects between the micro-ions. It
has repeatedly been demonstrated \cite{Groot,Levin} that in case of
monovalent micro-ions the PB theory provides a reliable description of
colloidal interactions. The interactions among colloids are, on this
level, mediated by the continuous distribution of the microions and
can be obtained once the local electrostatic potential due to the
microionic distribution is known. The normalized electrostatic
potential $\psi(x,y,z)$, which is the solution of the non-linear PB
equation,
\begin{eqnarray}
 \bnabla^{2} \psi(\vec{r}) & = & \kappa^{2} \sinh \psi(\vec{r})\;, 
\nonumber \\
{\bf n}\cdot\bnabla \psi & = & 4 \pi \lambda_{B} \sigma\;, \quad\quad
\vec{r} \;\rm{on}\, \rm{colloid}\, \rm{surface}\;,
\label{PBE}
\end{eqnarray}
describes the equilibrium distribution of the microions around a given
macroionic configuration. Here $\kappa$ is the inverse Debye screening
length , $\lambda_B$ the Bjerrum length ($\lambda_{B}=0.72$nm for
aqueous solutions at room temperature) and $\sigma$ is the surface
charge density on the colloid surface (constant charge boundaries are
assumed for all colloids in the system). ${\bf n}$ is the normal unit
vector on the colloid surface. We used the multi-centered technique,
described and tested in other studies \cite{CPC,EPL,JCP,JPCM} to solve the PB
equation (\ref{PBE}) at fixed configurations of three colloids and
obtain the electrostatic potential $\psi(x,y,z)$, which is related to
the micro ionic charge density. Integrating the stress tensor,
depending on $\psi(x,y,z)$, over a surface enclosing one particle,
results in the force acting on this particle. First, we calculated how
the force $f_{12}$, and from it the pair-potential between two
particles, depend on the distance between isolated two particles.
Choosing the suitable bare charge on the colloid surface, we were able
to reproduce the measured pair-interaction in Fig.\ref{pairpot}. The
calculation of three-body potentials was then carried out by
calculating the total force acting on one particle in the line trap
(say, particle 1) in the presence of all three particles and
subtracting the corresponding pair-forces $f_{12}$ and $f_{13}$
obtained previously in the two-particle calculation. If there is any
difference between the force on particle 1 obtained from the full PB
solution for the three particle configuration and the sum of two
two-body forces, this difference is due to the three-body interactions
in the system. The difference is then integrated to obtain the
three-body potential. The results are plotted as dashed and dotted
lines in Fig.\ref{u3} and show qualitative agreement with the
experimental data. To account for the deviations from the experimental
data one has to take into account the following points: {\bf (i)}
there is a limited experimental accuracy to which the light potential
can be determined. The accuracy decreases with increasing laser
intensity (note that normalized potentials are plotted in
Fig.\ref{laserpot}). In the three-body experiments, due to the
presence of the third repulsive particle, a stronger light field is
needed and the experimental error in determining the light potential
is estimated to be around $\pm 1 k_{B}T$. Since we have to subtract
the light potential twice from the total potential to obtain the
three-body potential, this error doubles and we expect an error of
about $\pm 2 k_{B}T$ in the final result. {\bf (ii)} An error of about
$\pm 2k_{B}T$ should be expected in the numerically obtained
three-body potentials as well. {\bf (iii)} While in the numerical
calculation we assume identical colloidal spheres, in the experiment
small differences with respect to the size and the surface charge are
unavoidable. This effect, however, is rather small and leads to
deviations on the order of 5 percent of the total potential.  {\bf
  (iv)} The numerical calculations do not take into account any
effects which may be caused by the substrate. Although we expect such
effects to be rather small (similar to the contribution for the pair
potential) they can not completely ruled out.  Considering the above
mentioned uncertainties it should be emphasized that in particular the
sign and the order of magnitude of the calculated potential compares
well with our measured results. This strongly supports our
interpretation of the experimental results in terms of the three-body
interactions.

Having solved the Poisson-Boltzmann equation we have the complete 
knowledge of the electrostatic potential in the space around the colloids
and can therefore observe the microscopic effects leading to the 
many-body interactions \cite{JCP}. The man-body interactions are a 
consequence of the nonlinearity of the physical equations governing the 
interactions in our system. Due to the nonlinearity the electrolyte density 
is different than a description based on the pairwise aditivity would 
predict. We have observed that there are less counterions in the region 
among the colloids than one would expect when superposing pair potentials. 
Less counterions means less screaning and larger electrostatic energy, but 
also larger entropy and at the end, obviously, smaller free energy. One 
could conclude, that on the microscopic electrolyte level the many-body 
interactions are entropy-driven.

\section{Conclusions}

We have demonstrated that the three-body interactions among charged 
colloids in suspension are attractive and of the same range as the 
pair-interactions. They present a considerable contribution to the
total interaction energy and have to be taken into account.
Further, for a limited number of configurations, also the four-body 
interactions have been explored. They have been found to be repulsive 
and of similar range and magnitude as three-body and pair terms. For the 
two configurations studied, the four-body interaction is almost precisely
the opposite of the three-body interaction in the system of three particles. 
Thus, adding the fourth particle and the four-body interaction term 
simply cancels one of the two three-body terms and the total many-body 
interaction potential $u_{\rm many-body}=2u_{123}+u_{1234}$ is equal to
the three-body term $u_{123}$ \cite{PRL,PRE}. The total many-body 
contribution stays attractive, although the four body interaction alone 
is repulsive.

Whenever dealing with systems comprised of many particles, in principle 
also higher-order terms have to be considered. The relative weight 
of such higher-order terms depends on the particle number density 
$\rho$. While at low enough $\rho$ a pure pair-wise
description should be sufficient, with increasing density first
three-body, four-body interactions and then higher-order terms 
come into play. One could speculate that there is an intermediate density regime, 
where the macroscopic properties of systems can be successfully described by
taking into account only two- and three-body interactions \cite{Anti} and, 
going up with the density, also the regime where two-, three- and four-body
terms are sufficient. Indeed liquid rare gases \cite{Jakse} and the island 
distribution of adsorbates on crystalline surfaces \cite{Osterlund} are 
examples where the thermodynamic properties are correctly captured by a 
description limited to pair- and three-body (AT-triple dipole interactions, 
either attractive or repilsive, depending on the configuration of the particles) 
interactions. 
 
However, it is difficult to predict in which range of parameters such an assumption
is justified. Especially the fact observed in this work, that the four-body
term compare with the three-body term in magnitude, should render such an approach
questionable. If one, describing a suspension, only takes two- and three-body 
effects into account, one takes too much attraction on board. We have seen in 
two examples in this paper, that the repulsive four-body interactions significantly 
reduce the attractive three-body interaction. Theoretical models taking two- and 
three-body effects into account \cite{Anti} are therefore of very limited value 
in describing a dense system. The continuation in this direction (five-, six- body, 
...interactions) does not seem promising, especially since the series even does not 
seem to converge and other methods have to be applied to capture the many-body effects. 
In the colloidal suspension all many-body interaction potentials are correctly taken 
into account by performing the non-linear Poisson-Boltzmann calculations. This has been 
done in  \cite{EPL,JCP,JPCM} where solid-liquid phase behaviour was investigated by 
Poisson-Boltzmann Brownian dynamics simulations. The conclusion there was that the total
many-body interaction does not vanish, but is attractive and leads to notable 
macroscopic effects, e.g. to a shift of the melting line in colloidal suspensions.


\begin{thebibliography}{99}

\bibitem{Louis} A.~A.~Louis, J.~Phys.: Condens.~Matter 14 (40), 9187 (2002). 
\bibitem{EPL} J.~Dobnikar, R.~Rzehak,
and H.~H.~von~Gr\"unberg, Europhys.~Lett. 61, 695 (2003).
\bibitem{JCP} J.~Dobnikar, Y.~Chen, R.~Rzehak, and
H.~H.~von~Gr\"unberg, J.~Chem.~Phys. 119 (9) (2003).
\bibitem{JPCM} J.~Dobnikar, Y.~Chen, R.~Rzehak, and
H.~H.~von~Gr\"unberg, J.~Phys.~Condens.~Matter 15(1) S263-S268 (2003).
\bibitem{gr} M.~Brunner, C.~Bechinger, W.~Strepp, V.~Lobaskin, and
  H.~H.~von~Gr\"unberg, Europhys.~Lett. 58, 926 (2002).
\bibitem{Teller} B.~M.~Axilrod and E.~Teller, J.~Chem.~Phys. 11, 299 (1943). 
\bibitem{Barker} J.~A.~Barker, D.~Henderson, Rev.~Mod.~Phys. 48, 587 (1976).
\bibitem{Bomont} J.~M.~Bomont and J.~L.~Bretonnet, Phys.~Rev.~B 65,
224203 (2002).  
\bibitem{Jakse} N.~Jakse,
J.~M.~Bomont, and J.~L.~Bretonnet, J.~Chem.~Phys. 116, 8504 (2002).
\bibitem{Formisano1} F.~Formisano, C.~J.~Benmore, U.~Bafile,
F.~Barocchi, P.~A.~Egelstaff, R.~Magli, and P.~Verkerk,
Phys.~Rev.~Lett. 79, 221 (1997).  
\bibitem{Formisano2} F.~Formisano,
F.~Barocchi, and R.~Magli, Phys.~Rev.~E 58, 2648 (1998).
\bibitem{Negele} J.~W.~Negele, Nucl.~Phys.~A 669, 18 (2002).
\bibitem{Hafner} J.~Hafner, From Hamiltonians to phase
diagrams (Springer, Berlin, Heidelberg, 1987).  
\bibitem{Osterlund}
L.~\"Osterlund, M.~O.~Pedersen, I.~Stensgaard, E.~Laegsgaard, and
F.~Besenbacher, Phys.~Rev.~Lett.~83, 4812 (1999).  
\bibitem{Ovchinnikov} M.~Ovchinnikov and V.~A.~Apkarian,
J.~Chem.~Phys. 110, 9842 (1999). 
\bibitem{Binder}
K.~Binder and D.~P.~Landau, Surf.~Sci. 108, 503 (1981). 
\bibitem{PRL} M.~Brunner, J.~Dobnikar, H.~H.~von~Gr\"unberg and C.~Bechinger,
Phys.~Rev.~Lett.~92, 78301 (2004).
\bibitem{PRE} J.~Dobnikar, M.~Brunner, H.~H.~von~Gr\"unberg and C.~Bechinger,
Phys.~Rev.~E, 69, 31402 (2004).
\bibitem{Palberg} T.~Palberg, W.~H\"artl, U.~Wittig, H.~Versmold,
M.~W\"urth, and E.~Simnacher, J.~Phys.~Chem. 96, 8180 (1992).
\bibitem{Landau} B.~V.~Derjaguin and L.~Landau, Acta Physicochim
U.R.S.S. 14, 633 (1941).  
\bibitem{Overbeek} E.~J.~W.~Vervey and
J.~T.~G.~Overbeek, Theory of the stability of lyophobic colloids
(Elsevier, Amsterdam, 1948).  
\bibitem{Belloni} L.~Belloni,
J.~Phys.~Cond.~Mat. 12, R549 (2000).  
\bibitem{Zsat} The saturated
effective charge of our particles is about 6900 for water at room
temperature.  
\bibitem{Trizac} E.~Trizac, J.~Phys.~A: Math.~Gen. 36, 5835 (2003).
\bibitem{Grier} S.~H.~Behrens and D. G. Grier, Phys.~Rev.~E 64, 50401
(2001).  
\bibitem{Stillinger} F.~H.~Stillinger, J.~Chem.~Phys. 35,
1584 (1961).  
\bibitem{Netz} R.~R.~Netz and H.~Orland,
Eur.~Phys.~J.~E. 1, 203 (2000).  
\bibitem{Burns} M.~M.~Burns, J.~M.~Fournier, and J.~A.~Golovchenko,
  Phys.~Rev.~Lett. 63, 1233 (1989).
\bibitem{McM} W.~McMillan and J.~Mayer,
J.~Chem.~Phys. 13, 276 (1945).   
\bibitem{CPC} J.~Dobnikar, D.~Halo\v zan, M.~Brumen, H.~H.~von~Gr\"unberg 
and R.~Rzehak, Comp.~Phys.~Comm., 159/2, 73-92 (2004).
\bibitem{Carsten}
C.~Russ, R.~van~Roij, M.~Dijkstra, and H.~H.~von~Gr\"unberg,
Phys.~Rev.~E 66, 011402 (2002).  
Europhys.~Lett. 58, 926 (2002).  
\bibitem{Groot} R.~D.~Groot,
J.~Chem.~Phys. 95, 9191 (1991).  
\bibitem{Levin} Y.~Levin,
Rep.~Prog.~Phys. 65, 1577 (2002). 
\bibitem{Anti} A.-P.~Hynninen, M.~Dijkstra and R.~van~Roij,
  J.~Phys.~Cond.~Mat. 15 (48), S3549 (2003).
\end{thebibliography}
\end{document}